\documentclass[pre,aps,preprint,showpacs,floatfix,superscriptaddress]{revtex4-1}

\usepackage {graphicx,amssymb}
\graphicspath{{fig/}}

\newcommand{\gae}
{\,\hbox{\lower0.5ex\hbox{$\sim$}\llap{\raise0.5ex\hbox{$>$}}}\,}
\newcommand{\lae} 
{\,\hbox{\lower0.5ex\hbox{$\sim$}\llap{\raise0.5ex\hbox{$<$}}}\,}
\newcommand{\be}{\begin{equation}}
\newcommand{\ee}{\end{equation}}
\newcommand{\bea}{\begin{eqnarray}}
\newcommand{\eea}{\end{eqnarray}} 
\newcommand{\bdm}{\begin{displaymath}}
\newcommand{\edm}{\end{displaymath}}

\begin{document}
\title{Critical manifold of the Potts model:\\
Exact results and homogeneity approximation}
 \author{ F. Y. Wu}
 \email{fywu@neu.edu}
\affiliation{Department of Physics, Northeastern University, Boston, 
Massachusetts 02115, USA }
 \author{Wenan Guo}
\email{waguo@bnu.edu.cn}
\affiliation{Physics Department, Beijing Normal University,
Beijing 100875, P. R. China}

\begin{abstract}
 The $q$-state Potts model has stood at the frontier of research in statistical 
mechanics for many years. In the absence of a closed-form
solution, much of the past efforts have focused on locating its critical
manifold, trajectory in the parameter  $\{q, e^J\}$  space where
$J$ is the reduced interaction, along which the free energy is singular.
However,  except in  
isolated  cases,  antiferromagnetic (AF)
  models with $J<0$ have been largely neglected. 
In this paper we consider the Potts model with AF interactions focusing on obtaining
 its critical manifold in  exact and/or closed-form expressions.
  
We first re-examine the known critical frontiers
  in light of 
AF interactions. For the square lattice we confirm  the Potts self-dual point to be
the sole critical frontier for $J>0$.  We also locate its 
critical frontier for $J<0$ and find it to  coincide with a solvability condition observed by Baxter in 1982.
  For the 
honeycomb lattice we show that the known critical frontier   holds for {all} $J$, and determine
 its  critical
 $q_c = \frac 1 2 (3+\sqrt 5) = 2.61803$ beyond which there is no transition. 
For the triangular lattice we confirm the known critical frontier  to hold only for $J>0$.

More generally we consider the centered-triangle (CT) and Union-Jack (UJ) lattices 
consisting of mixed  $J$ and $K$ interactions, and deduce  critical manifolds under 
  homogeneity hypotheses.  For $K=0$ the CT lattice is
 the diced lattice, and we determine its critical manifold for all $J$ and find
 $q_c = 3.32472$.
 For $K=0$ the  UJ lattice is the square lattice and from this we 
deduce both the $J>0$ and $J<0$ 
critical manifolds  and  $q_c=3$.
 Our theoretical
 predictions  are compared with recent   numerical results.  
\end{abstract}
\maketitle

The Potts model, proposed by Potts exactly sixty years ago \cite{Potts}, has been at the
 forefront of  interest throughout the years.  The Potts model extends
 the 2-state Ising model to  $q >2$ states (for  reviews on relevances of the Potts model 
see  \cite{WuPotts,WuPotts1}).    Despite  the intense interest, however,
the general $q$-state Potts model   has remained unsolved.   
 An equivalent and more revealing formulation of the Potts model is the 
{random cluster} model advanced by Kasteleyn and Fortuin \cite{FK2,FK1}.  
  In  this formulation
 the partition function is written as a sum of randomly connected cluster of
lattice sites with edge weights $v=e^J -1$,
 where $J$ is the  
Potts interaction with $J>0$ (resp. $<0$) indicating ferromagnetic (resp. antiferromagnetic (AF)) interactions.  
The number of spin states $q$ then emerges as a cluster weighting factor,  
 thus providing a powerful means of continuing the Potts model to non-integral values of $q$.
  In the limit of $q\to 1$, for example,
 the formulation leads to a bond percolation with edge occupation probability
 $p = 1 - e^{-J}$. 
 
Most studies of the Potts model 
 have focused on  ferromagnetic interactions, and the AF models have been largely
neglected. 
 (See however \cite{JS,IJS} where the AF Potts model is discussed in the context of field theory.)
For $J<0$, 
the cluster weights which are products of edge weights $v$, 
 can be either
positive or negative resulting in a delicate cancelation. 
  Since most
analyses are based on a positivity assumption
of  Boltzmann weights, they
are  not applicable to   AF models.  
In the percolation limit, for example,
 the edge occupying probability  becomes negative losing its physical meaning.

In the absence of closed-form solutions, efforts have focused on
 determining its critical  manifold, or critical frontier,  a trajectory in the $\{ q, e^J \}$
parameter space along which the free energy becomes singular. But known critical manifold of the Potts model
are very few.  They are limited   to the square, honeycomb
 and triangular
lattices given by
\bea
v^2 &=& q \>\>\>({\rm square})  \label{square} \\
v^3-3qv - q^2 &=& 0\>\>\>({\rm honeycomb})  \label{hc} \\
v^3 + 3v^2 - q &=& 0. \>\>\>({\rm triangular})  \label{tri}
\eea
Here, we examine closely the applicability of (\ref{square}) - (\ref{tri})
 to  AF interactions.

 The critical point (\ref{square}) for the square lattice was obtained by Potts 
\cite{Potts}    using the duality relation \cite{WuPotts}
\be
(e^J-1)(e^{J^*} -1) = q  \label{sq}
\ee
 relating an interaction $J$ to a $J^*$ on the dual lattice. Since the square lattice
is self-dual and  assuming a unique transition,  the duality (\ref{sq}) determines the critical point  
at $J = J^* $ or $v^2=q$.  This works only for $J>0$.  
For  $J<0$ and $q>1$, (\ref{sq}) relates $J$ to a dual $J^*$ 
in the complex-temperature plane. The self-dual argument  fails and 
 (\ref{square}) does not hold.
 
In 1982,  Baxter \cite{Baxter1982}  observed that the AF Potts model on the square lattice is exactly solvable at
\be
(e^J+1)^2 = 4 -  q,  \quad\quad (q \leq 3)  \label{AFsqcritical}
\ee
and argued that the condition (\ref{AFsqcritical})
should coincide with the 
critical point as in the case of the ferromagnetic model. 
Indeed, as we shall see below, 
the condition (\ref{AFsqcritical})  
emerges as the AF branch of the Potts critical manifold when the square lattice is
considered
in the larger setting of a Union-Jack lattice.

One important feature of the 
$J<0$  transition is the existence of a critical 
$q_c$ beyond which there is no  transition. 
  The disappearance of a transition for large $q$ in any AF model
is expected on general ground since for $q$ sufficiently large 
the ground state entropy  will destroy any transition. But the actual determination of
$q_c$ is lattice-dependent. For the square lattice, for example, (\ref{AFsqcritical})
gives $q_c=3$. 
 
The critical points (\ref{hc}) and (\ref{tri}) for the honeycomb and triangular lattices can
be derived as follows:  Consider a triangular Potts lattice whose 
Hamiltonian is separable into sum over individual up-pointing
triangles so that
  the partition function can be written as
\be
Z= \sum_{s_i=1}^q  \prod_{\bigtriangleup} W(s_1, s_2, s_3), \label{Z}
\ee
where $W(s_1, s_2, s_3)$ is the Boltzmann weight of an individual triangle with 3
corner spins $s_1, s_2, s_3$;  the product
is taken over all up-pointing triangles. An example is  the honeycomb lattice
divided into  triangles shown in Fig.\ref{st1}(a).
   Clearly,
   properties of this Potts model are completely determined
by the weights $W(s_1, s_2, s_3)$.

Baxter {\it et al} \cite{Baxter1978}  (see also  Wu and Lin \cite{wulin}) have obtained a self-dual
trajectory for a triangular Potts model with 2- and 3-site interactions. 
In terms of the $W$-weights the self-dual trajectory
can be recast as
\be 
W(1,1,1)= 3\,  W(1,1,2) + (q-2)\, W(1,2,3). \label{hcgeneral}
\ee
 Wu and Zia \cite{wuzia} have further established that, 
for positive $W(s_1, s_2, s_3 )$ and in the regime $W(1,1,1)$ dominates, or 
 \be
W(1,1,1) > \big\{ W(1,1,2),\> W(1,2,3) \big \} >0 , \label{regime}
\ee
there is a transition into a ferromagnetic ordered state 
on the underlying triangular lattice at the  self-dual point (\ref{hcgeneral}).
Indeed, for $W>0$ the self-dual trajectory (\ref{hcgeneral}) lies entirely in the 
regime (\ref{regime}). 
The transition is expected to be continuous
for $q\leq 4$ and of first order for $q>4$ from universality considerations \cite{WuPotts}.
In the regime where $W(1,2,3)$ dominates and for $q=3$, there exists a transition 
into an antiferromagnetic ordered state  which has  a 6-fold degenerate ground state. 
But the critical manifold for this transition in terms of the $W$-weights is not known.
  
We now specialize ({\ref{hcgeneral}) to specific lattices.
For the honeycomb  Potts lattice we take as the individual up-pointing triangles  the setup
in Fig. \ref{st1}(a). This gives after  tracing over the center spin $s_0$,
 \be
W(1,1,1)= e^{3J}+q-1, \
W(1,1,2)=  e^{2J}+ e^{J}+q-2, \
W(1,2,3)= 3 e^{J}+q-3.  \label{newhcweights}
\ee
 Substituting (\ref{newhcweights}) into (\ref{hcgeneral}) we obtain the critical point (\ref{hc}).  
  Since $W\geq 0$ for all (positive and negative) $J$, 
 the critical manifold (\ref{hc}) holds for {\it all} $J$ and gives
the critical $q_c$ as the root of
\be
q^2 - 3q +1 = 0, \quad\quad ({\rm honeycomb}) \nonumber
\ee
or $q_c  = \frac 1 2 (3+\sqrt 5) = 
2.61803... $.

For latter use we 
recast the critical manifold (\ref{hcgeneral}) in another language.
The star-triangle transformation of Fig. \ref{st1}
transforms the honeycomb lattice 
 to a triangular lattice with 2-site interactions $K'$ and 3-site
interactions $M$ in every up-pointing triangle. A little algebra shows
that the critical manifold (\ref{hcgeneral}) can be written as
\be
e^{3K'+M}   -3\  e^{K'} - (q-2) = 0, \label{trian}
\ee
with
 \be
e^{3K'+M} = \frac {e^{3J}+q-1} {3e^{J}+q-3} , \quad
 e^{K'} = \frac {  e^{2J}+{e}^{J}+q-2} {  3e^{J}+q-3}.
\label{MK}
\ee

Now turning  to the triangular Potts lattice for which each individual  triangle
consists of 3 edges of interactions $J$, we have
\be
W(1,1,1)= e^{3J}, \quad \quad
W(1,1,2)=  e^{J}, \quad \quad
W(1,2,3) =1 . \label{newtriweights}
\ee
For $J>0$ the weights (\ref{newtriweights}) are in the regime (\ref{regime}) and
the substitution of (\ref{newtriweights}) into (\ref{hcgeneral}) leads to the critical 
manifold (\ref{tri}). For $J<0$ and $q=3$, the weights (\ref{newtriweights}) are in the AF regime.
  The location of its critical point 
 has been the subject matter of several studies. 
From 
low-temperature series of the partition function Enting and Wu \cite{enting} determined 
a  first-order transition at the critical point $e^{J_c} = 0.204$. Adler {\it et al} \cite{Adler}
found $e^{J_c} = 0.20309 \pm 0.00003 $ from
Monte Carlo simulations, and Chang et al \cite{Chang} found
$e^{J_c} =0.203073(20) $ from transfer-matrix computations.
 Feldmann {\it et al} \cite{feldmann} 
showed this transition  to be consistent with a complex
singularity in the dual honeycomb lattice \cite{jensen}.  
  
\bigskip
\begin{figure}[htbp]
\includegraphics[scale=0.4]{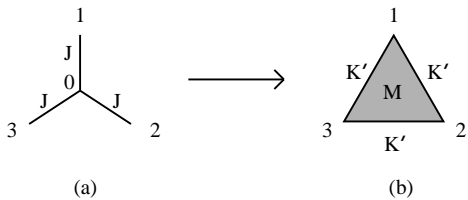}
\caption{A star-triangle transformation. The 'star' in (a) is mapped into
a triangle in (b) consisting of 2-spin interactions $K'$ and 3-spin interaction $M$.}
\label{st1}
\end{figure}

We consider next the critical point of the diced lattice, a subject matter of recent studies
  \cite{kotecky2,Chen}.  
The diced lattice  is the lattice in Fig. \ref{UJL}(a) with the heavy edges $K$ removed. 
 In 1979, one of us \cite{WuJPC} (see also \cite{WuPRE}) introduced a homogeneity hypothesis 
and  derived
a  conjectured 
critical manifold for the diced lattice. The derivation of the homogeneity hypothesis
is as follows:

Apply the star-triangle transformation
 of Fig. \ref{st1} to {all} 3-coordinated sites of the diced lattice. This 
results in a triangular
lattice with 2-site interactions $2K'$ and 3-site interactions $M$ in {\it every} face
with $e^{K'} $ and $e^{M}$ given by (\ref{MK}). 
If the interaction $M$ exists in only the up-pointing triangular faces, the exact critical frontier
is (\ref{trian}) with $K'\to 2K'$.  
  The idea of homogeneity hypothesis, which we shall later extend to 
the Union-Jack lattice, is to postulate that when 3-site interactions $M$ and $M'$ are present in
alternating faces, the critical frontier is (\ref{trian}) modified by  replacing $M \to M+M'$ in 
the exponential.
In the case of diced lattice where $M'=M$, 
 this gives the critical manifold (\ref{trian}) 
with   $K'\to 2K', M\to 2M$.  This is the 
homogeneity approximation (HA).  

The resulting critical manifold
  can  be written in a generic form
\be 
W^2(1,1,1)=3\, W^2(1,1,2)+ (q-2)\ W^2(1,2,3) \quad \quad \quad \quad \quad \quad ({\rm HA}) \label{HAA}
\ee
with $W$-weights  given in (\ref{newhcweights}). 
Explicitly,  (\ref{HAA}) reads
 \be
v^6 + 6v^5 +12 v^4 +2q v^3 - 9 qv^2 -6 q^2 v - q^3  =0.
\quad \quad({\rm HA - diced\>\>lattice}) \label{diced}
\ee
The critical manifold for the kagome lattice is  the dual
 of (\ref{diced}), or
\be
v^6 +6v^5 +9v^4 -2 q v^3 - 12 q v^2 - 6q^2 v -q^3 = 0. 
\quad \quad ({\rm HA - kagome\>\> lattice}) \label{kagome}
\ee
This is the conjecture  advanced in \cite{WuJPC}.
In the case of $M\not= M'$ and the up- and down-pointing triangular faces have different weights
$W_\bigtriangleup$ and $W_\bigtriangledown$, the generalization of (\ref{HAA}) is to
replace $W^2$ by $W_\bigtriangleup\cdot  W_\bigtriangledown$.

The critical manifolds (\ref{diced}) and (\ref{kagome})  are exact for
  the Ising model $q=2$.    At $q=2$  (\ref{diced}) gives the solution
 $ \cosh J = (\sqrt 3 +1 )/2$ and  (\ref{kagome}) 
    has only one   solution at $e^J = \sqrt {3+2\sqrt 3} = 2.5424598$. both 
in agreement with Syozi \cite{Syozi}.

At $q=3$, 
(\ref{diced}) has six roots, $(e^J)_{\rm diced} = v+1 $  $= 2.598917,$ $0.121599,$ $ -2.255914,$ $ -0.462026,$  
and $-0.00128784 \pm 1.742380\  i$.  The first 2 roots give 
 respective  $J>0$  and $J<0$ critical points.
  The other 4 solutions are outside the regime (\ref{regime}) and  are not singular points, 
so they are discarded.
The $J>0$ transition point $ e^J = 2.598917$ is within $0.002\%$ of the high-precision 
value 
 $e^J = 2.598755(6)$ from a preliminary finite-size analysis \cite{Fu}.
But the $J<0$ transition point $0.121599$ 
differs by about $8\%$ from the value
 $0.1393650(4)$  from the same 
 finite-size analysis. 
Earlier numerical estimates include 
$  0.1394$ and $  0.1380$ of 
 Kotecky {\it et al} \cite{kotecky2} and Chen {\it et al}  \cite{Chen}, 
and the value
$0.1393(8)$ of Feldmann {\it et al} \cite{Feldmann1} 
computed from a complex-temperature singularity of the dual kagome lattice \cite{jensen}.
All these values are consistent with the precise determination $0.139365$ of \cite{Fu}.
The relatively sizable discrepancy of the HA prediction indicates its
inadequacy for AF interactions.

At
$v=-1$  (\ref{diced}) determines $q_c$ for the diced lattice from 
 \be
 q^3-6q^2+11 q -7 =0, \quad\quad ({\rm diced}) 
\ee
or $q_c=3.32472$.  This is compared to the bound $q_c= 3.117689$ argued by Kotecky {\it et al}
\cite{kotecky2} and the more precise value $q_c = 3.4380(2)$ from finite-size analysis \cite{Fu}.

The critical points of the kagome lattice are the dual of those of the diced lattice.
For $q=3$ the singularity at  $ 2.598917$ of the diced gives 
a corresponding kagome critical point
$ 2.876 269 $  which has been shown  \cite{Ding} to be of the same high accuracy as the diced solution.
The dual of the  AF  singularity is complex having no physical meaning.
 So there is only a single (ferromagnetic) transition in the kagome lattice.
Note that a
 direct  evaluation of (\ref{kagome}) yields a spurious AF  solution $e^{J_c} = 0.078600$
which is the dual of the discarded  diced solution $-2.255914$ not representing a singularity.

 \begin{figure}[htbp]
 \includegraphics[scale=0.22]{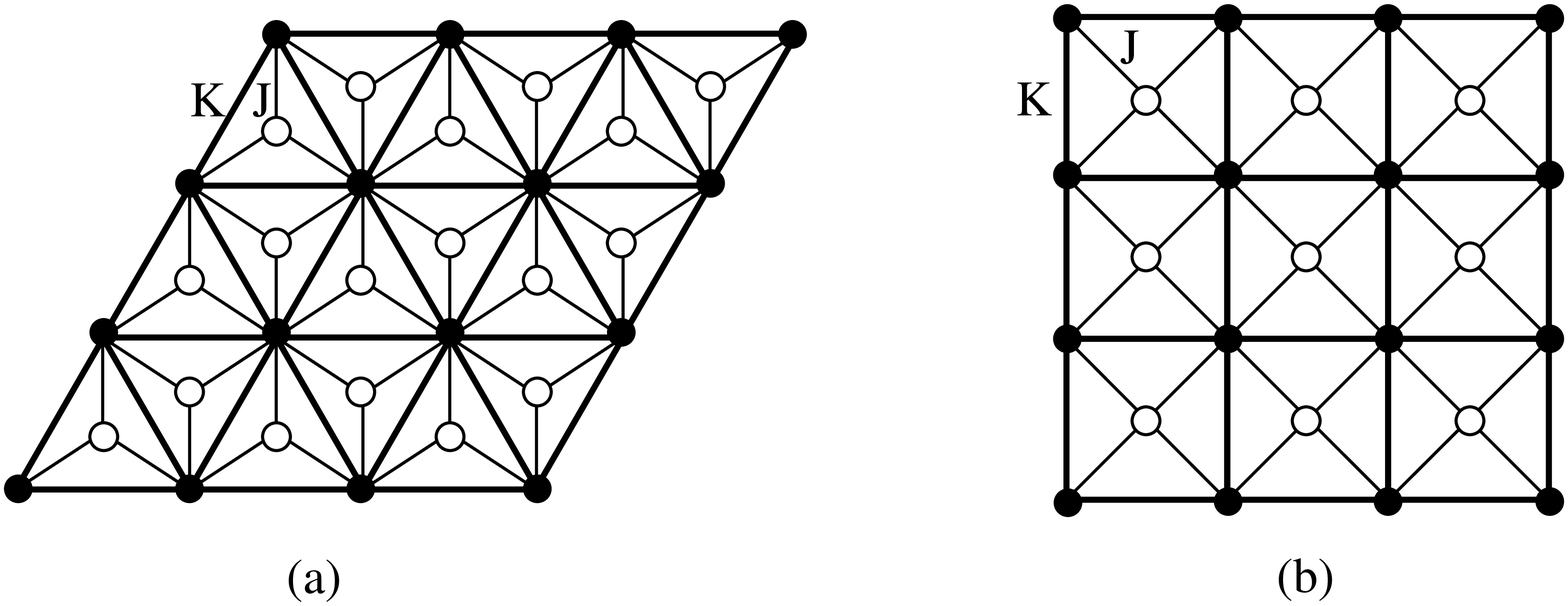}
\caption{(a) The centered-triangle lattice 
constructed by inserting sites (open circles) to
faces of the triangular lattice.  (b) The Union-Jack lattice.} 
 \label{UJL}
\end{figure}

We consider next the centered-triangular (CT) and Union-Jack (UJ) lattices shown in
Fig. \ref{UJL}. The CT lattice, also known as the 
asanoha or hemp-leaf lattice \cite{Syozi},
is a triangular lattice with a spin inserted in each triangular face.
Similarly the UJ lattice can be regarded as a centered-square lattice.
These  lattices have received considerable recent attention as instances of 
finite-temperature transitions in AF Potts models
\cite{kotecky2,Chen,Deng}. 
 
For the CT lattice we trace over all 3-coordinated sites.  This leads  
to a triangular lattice with 2-site 
interactions $K+2K'$ and 3-site interactions $M$ in every face. Then, as in the case
of the diced lattice, the 
homogeneity  CT critical manifold is 
(\ref{trian}) 
with $K'\to K+2K', M\to 2M$. This leads to the
critical manifold (\ref{HAA}) 
with 
 \bea
W(1,1,1) &=& e^{3K/2}(e^{3J}+q-1),\  \nonumber\\
W(1,1,2)  &=& e^{K/2}(e^{2J} +e^J +q-2),\ \nonumber \\
 W(1,2,3) &=&  3 e^{J}+q-3.\label{HAB}
\eea
(Note the split of the interaction  $K$ into two parallel $K/2$.) Explicitly, the critical manifold reads
\be
e^{3K}(e^{3J}+q-1)^2 = 3\ e^K (e^{2J} +e^J +q-2)^2 
  + (q-2) (3 e^{J}+q-3)^2 ,\quad  ({\rm HA - CT \>\>lattice}) \label{HAC}
\ee
which is to hold in  regime (\ref{regime}) with $W$-weights (\ref{HAB}).

The critical manifold 
({\ref{HAC}) is exact for  the Ising model since
the star-triangle transformation is exact
for $q=2$.  This gives \ $ 2 \cosh J  = 1 + \sqrt 3 \ e^{-K}$. 
In the general case we have considered the solution of  (\ref{HAC}) for $|K| = |J|$.
 We find  $q=3$ solutions   $e^J = e^K = 1.596063 $ and  $e^J = e^{-K} = 0.602542$.
 These  values are accurate to within 5 decimal places when compared with 
 results of a preliminary
finite-size analysis  \cite{Fu}. There is no $K <0, |J| = |K|$ solution.  
 
The Union-Jack is the lattice shown in Fig. \ref{UJL}(b).
 In order to extend  the homogeneity consideration, 
we consider first the 
anisotropic UJ lattice with a unit cell shown in Fig. \ref{uj}(a). Setting $J_4 = 0$ as   in Fig. \ref{uj}(b),
 the UJ lattice becomes the triangular lattice   in Fig. \ref{uj2tri} 
with  $J_1, J_2, J_3$ interactions in 
up-pointing triangular faces.  
The critical manifold of this  triangular lattice 
is  the anisotropic version of (\ref{hcgeneral}) \cite{wuzia}, namely.
 \be
W(1,1,1) =  W(1,2,2)+ W(2,1,2) 
 + W(2,2,1) \ + (q-2)\ W(1,2,3).   \label{hcgeneral1}
\ee
Here  (\ref{hcgeneral1}) is to hold 
 in the regime $W>0$ and $W(1,1,1)$ dominates, and 
 the transition is to a state of ferromagnetic ordering.

The $W$-weights in (\ref{hcgeneral1}) can
be read off from Fig. \ref{uj}(b),
\bea
&& W(1,1,1)= e^{K_1+K_2}(e^{J_1+J_2+J_3}+q-1),\nonumber \\
&&  W(1,2,2)=e^{K_2} (e^{J_1}+e^{J_2+J_3}+q-2), \nonumber \\
&& W(2,1,2)=e^{J_2}+e^{J_1+J_3}+q-2, \nonumber \\
&& W(2,2,1)=e^{K_1}(e^{J_3}+e^{J_1+J_2}+q-2), \nonumber \\
&& W(1,2,3)=e^{J_1}+e^{J_2}+e^{J_3}+q-3, \label{WCT}
\eea
and the critical manifold (\ref{hcgeneral1}) with (\ref{WCT}) is exact. 

 \begin{figure}[htbp]
\includegraphics[scale=0.23]{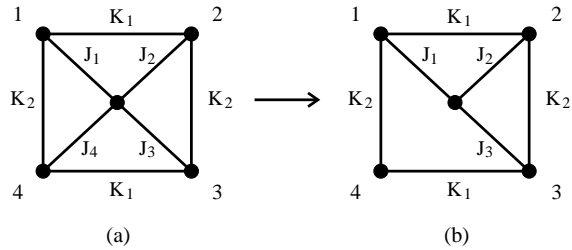}
\caption{(a) Union-Jack lattice with anisotropic interactions.
  (b) The $J_4=0$ Union-Jack lattice. }
\label{uj}
\end{figure}
\begin{figure}[htbp]
\includegraphics[scale=0.35]{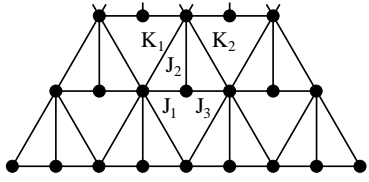}
\caption{A triangular lattice with $J$ interactions in every up-pointing triangular face. }
\label{uj2tri}
\end{figure}

Now consider the Union-Jack lattice with $J_4 \not= 0$.  
As in the case of the CT lattice, we introduce the homogeneity hypothesis 
by inserting $J_4$ into (\ref{hcgeneral1}) such that the resulting expression retains appropriate
symmetries of the 4 interactions $J_1, J_2, J_3, J_4$, and reduces to (\ref{hcgeneral1}) 
 when $J_4$  is set
to zero. 
The simplest way to do this is to insert  $J_4$ into appropriate
exponents in $W(1,1,1),W(1,2, 2), W(2, 1, 2), W(2, 2, 1)$
and insert 
$e^{J_4} - 1$ into $W(1,2,3)$.  This gives rise to
 the following homogeneity approximation for the anisotropic UJ lattice,
\bea
e^{K_1+K_2} (e^{J_1+J_2+J_3+J_4} +q-1) 
&=&  e^{K_2}(e^{J_1+J_4}+e^{J_2+J_3} +q-2) \nonumber \\
&+& e^{K_1}(e^{J_3+J_4}+e^{J_1+J_2} +q-2) 
+ e^{J_2+J_4}+e^{J_1+J_3} +q-2 \nonumber \\
&+& (q-2)(e^{J_1}+e^{J_2}+e^{J_3}+e^{J_4} +q-4).
\label{ujconj1}
\eea
When $K_1=K_2=0, $ (\ref{ujconj1}) reduces to the critical manifold of the checkerboard
lattice conjectured in \cite{WuJPC}.  However, when $K_1 \ne K_2$ the expression (\ref{ujconj1})
lacks the  intrinsic  symmetry 
of $J_1 \leftrightarrow J_3$ or $J_2\leftrightarrow J_4$.  For the isotropic lattice
  $J_1=J_2=J_3=J_4=J$ we are interested in, we obtain
\bea
e^{K_1+K_2} (e^{4 J} +q-1) 
= (e^{K_1}+e^{K_2}) (2 e^{2 J}+q-2) &+&  2\ e^{2J} +  4(q-2) e^{J} 
+(q-2)(q-3).\>\  \nonumber \\
&&({\rm HA - Union\>\>Jack\>\>lattice})
\label{ujconj2}
\eea 
This is the homogeneity approximation for the Union-Jack lattice.
  
 The manifolds (\ref{ujconj1}) and (\ref{ujconj2}) are exact at $q=2$ as they reproduce the 
known Ising critical point obtained by Wu and Lin \cite{wulin87}.
When $J=0$, the Union-Jack lattice reduces to a square
lattice with anisotropic interactions $K_1$ and $K_2$, and
 (\ref{ujconj2}) gives the known critical point 
$ (e^{K_1}-1)(e^{K_2}-1)=q$. 
When $K_1 =K_2 =0$, the Union-Jack lattice reduces to a square lattice 
of interactions $J$ and (\ref{ujconj2}) factorizes into
 \be
 \ \big[(e^J-1)^2 -q\big] \ \big[e^{2J} +2 e^J +q -3\big] =0.
\ee
 Setting the first factor equal to zero gives the critical frontier (\ref{square}).
Setting the second factor equal to zero gives the  Baxter solvability condition 
 (\ref{AFsqcritical}) as the $J<0$ critical frontier as alluded to earlier.

In the general case we have considered the solution of  (\ref{ujconj2})   for $|K| = |J|$.
For $q=3$ we found solutions  $e^J = e^K = 1.80168 $ and $e^J = e^{-K} = 0.49256$. 
In addition, there is  
 an AF solution   in $2\leq q \leq 3$ at, 
for example, $e^K= e^J = 0.081504 $ at $q=2.5$.  This solution is 
 an extension of the $J<0$ transition (\ref{AFsqcritical}). 
For $K_1=K_2=K$ and setting $e^K = e^J = 0$ in (\ref{ujconj2}), we obtain $q_c=3$.  Indeed,
numerically we found
no $K=J<0$ solution of (\ref{ujconj2})
in $q>3$.
  We remark that Chen {\it et al} \cite{Chen} have recently studied the $q=4$, $K=J <0$, 
Union-Jack model (see also \cite{Deng}) using a tensor-based numerical method, 
and  found an 'entropy-driven'  transition 
at  $e^K=e^J = 0.0523$. Since we found $q_c=3$  for transitions to 
ferromagnetic ordering on the underlying square lattice, 
  the transition 
at $q =4$  is likely of a  different nature
 as described and argued by Deng et al \cite{Deng}.
  
Finally we remark that since the duals of the CT and UJ lattices are, respectively, the 3-12 and 4-8, 
or ($3, \ 12^2$) and ($4,\, 8^2$), 
 lattices,  the duals of (\ref{HAC}) and
(\ref{ujconj2}) give the critical manifolds for the 3-12 and 4-8 lattices.  
In the case of  uniform interactions $K=K_1=K_2=J$, 
we obtain
\bea
v^7(v+3)^2 &=& q\big[3v^7 + (q+32)v^6 +15(2q+5)v^5  
+ q(12q+ 111)v^4 +2q^2(q+41)v^3 \nonumber \\
&+& 36q^3v^2 +9q^4v +q^5\big],
\quad \quad \quad \quad \quad \quad \quad \quad \quad \quad ({\rm 3-12\>\>lattice}) \label{3-12} \\ 
 v^6+4v^5 &=& q\ (v^4+16v^3+15qv^2+6q^2v+q^3),
  \qquad \qquad({\rm 4-8\>\>lattice}) \label{4-8}
\eea
where $v=e^J-1$.
Both (\ref{3-12}) and (\ref{4-8}) 
are exact at $q=2$ reproducing the known Ising critical points \cite{Syozi}, and have
only  ferromagnetic solutions.
 At $q=1$ 
(\ref{3-12}) and (\ref{4-8})  give the bond percolation thresholds $\big(p_c\big)_{3-12}=0.740423$ and
$\big(p_c\big)_{4-8}=0.676835$ in agreement with prior numerical estimations \cite{percolation}.
 
In summary, we have considered a lingering unsettled question on critical manifolds of
 Potts models with AF interactions.
We have re-examined the known critical frontiers
of the square, honeycomb  and the triangular lattices in light of AF interactions.
We have also examined the critical manifolds of the  diced 
and kagome lattices under a homogeneity assumption, and
extended the homogeneity consideration to  centered-triangular and Union-Jack lattices
to deduce  critical manifolds.
Our theoretical predictions are compared with results of several recent numerical studies.

After the completion of this work, we 
received a preprint \cite{job} 
on a new
graphical analysis of  the Potts  critical manifold
which identifies the kagome conjecture (\ref{kagome}) 
as a  first-order approximant, thus resolving the long-standing
 question on the exactness of the conjecture.
 In \cite{job} the AF critical line (\ref{AFsqcritical})
is also deduced from the checkerboard manifold.
We thank J. L. Jacobsen for  sending a copy of \cite{job} prior to publication. 
The identification of the homogeneity approximation
as a first-order approximant in the graphical analysis can also be extended  to
(\ref{ujconj2}) for  $K_1=K_2$  \cite{job1}.
 We would like to
thank R. Shrock for comments and suggestions and calling attention to \cite{feldmann,jensen,Chang},
and thank J. Salas and R. Ziff for comments. 
We thank Z. Fu for help in the preparation of the
manuscript. The work of WG is supported by the National Science Foundation of China (NSFC) under Grant No. 11175018, 
and
by the program for New Century Excellent Talents in University (NCET-08-0053).

\end{document}